# Josephson effect between electron-doped and hole-doped iron pnictide single crystals

Xiaohang Zhang,[a)] Shanta Saha, Nicholas P. Butch, Kevin Kirshenbaum, Johnpierre Paglione, and Richard L. Greene

*CNAM and Department of Physics, University of Maryland, College Park, Maryland 20742, USA*

Yong Liu, Liqin Yan, Yoon Seok Oh, and Kee Hoon Kim

*CSCMR & FPRD, Department of Physics and Astronomy, Seoul National University, Seoul 151-747, Republic of Korea*

Ichiro Takeuchi

*Department of Materials Science and Engineering, University of Maryland, College Park, Maryland 20742, USA*

(Dated on June 9, 2009)

**ABSTRACT** We have observed the Josephson effect in junctions formed between single crystals of $SrFe_{1.74}Co_{0.26}As_2$ and $Ba_{0.23}K_{0.77}Fe_2As_2$. *I-V* curves showed resistively-shunted junction characteristics, and the ac Josephson effect was observed under microwave irradiation. By applying an in-plane magnetic field, the critical current is completely modulated and shows a relatively symmetric diffraction pattern, consistent with the intermediate junction limit. The observation of the Josephson effect in the *p-n* bicrystal structure not only has significant implications for designing phase-sensitive junctions to probe the pairing symmetry of iron pnictide superconductors, but also represents an important step in developing all iron pnictide devices for applications.

a) Email address: xhzhang@umd.edu

Previously, we have demonstrated Josephson coupling between a single crystal pnictide superconductor ($Ba_{1-x}K_xFe_2As_2$) and a conventional superconductor along the *c*-axis, which indicated the presence of an *s*-wave order parameter in the iron pnictide superconductor.[1] In this letter, we report the observation of the Josephson effect in bicrystal junctions along the *c*-axis between electron-doped (e-doped) and hole-doped (h-doped) pnictide single crystals. The critical currents are completely modulated by applying an in-plane external magnetic field. The obtained magnetic diffraction patterns are Fraunhofer-like with well defined modulation periods.

Fabrication of such *p-n* junctions is motivated by a number of reasons: 1) there have been a number of proposals[2-4] of definitive phase-sensitive tests for the widely discussed *s±*-wave symmetry[5-8] in pnictide superconductors which involve transport across pnictide *p-n* interfaces; 2) as in the case of high-$T_C$ cuprates,[9] the presence of both electron and hole doping in the pnictides raises an intriguing question about the possibility of forming rectifying *p-n* junctions; 3) such junctions represent an important step towards making all-pnictide Josephson devices.

The e-doped ($SrFe_{1.74}Co_{0.26}As_2$) and the h-doped ($Ba_{0.23}K_{0.77}Fe_2As_2$) iron pnictide single crystals used in this study were grown in FeAs flux[10] and Sn flux,[11] respectively. The single crystals were in platelet shapes with the normal direction along the *c*-axis. Wavelength dispersive X-ray spectroscopy (WDX) and energy dispersive X-ray spectroscopy (EDX) were used to determine the doping concentrations in the e-doped and the h-doped single crystals. Magnetic susceptibility measurements confirmed bulk superconductivity in both types of single crystals. Resistivity measurements (Fig. 1a) showed the superconducting transitions were very sharp, and the $T_C$'s were approximately 17 K for $SrFe_{1.74}Co_{0.26}As_2$ and 22 K for $Ba_{0.23}K_{0.77}Fe_2As_2$. The Hall effect measurements (Fig. 1b) showed that the Hall coefficient of each type of crystal had a clear temperature dependence with no sign change. The negative (positive) sign of the Hall coefficient of $SrFe_{1.74}Co_{0.26}As_2$ ($Ba_{0.23}K_{0.77}Fe_2As_2$) single crystals confirms that the transport is dominated by electrons (holes), which is consistent with the dopant type of the compound and in qualitative agreement with previous reports.[12,13] In both types of single crystals, all Hall resistivity curves display a strong linear dependence in magnetic fields up to 7 T. As an example, Fig. 1c shows the Hall resistivity curves obtained at 35 K.

Scanning electron microscopy (SEM) of the crystals indicated that their surfaces consisted of large terrace-free areas up to mm$^2$ in size.[1] Crystals were cut into rectangular pieces with a typical size of ~ 1500 μm × 300 μm along *a* or *b* axis edges. Each crystal was soldered on a sapphire substrate by indium, and two separated electrical contacts were made on the back of the crystals. The e-doped and the h-doped single crystals were pre-aligned in a cross geometry as illustrated in the inset of Fig. 2a, so that the *a* and the *b* axes are coincident (parallel or orthogonal) for the two crystals. At the liquid helium temperature, the two crystals were then pressed together to produce a bicrystal junction with an intended overlap junction area of ~ 300 μm × 300 μm. We have studied several vertically aligned bicrystal junctions fabricated in this manner, and the junctions showed qualitatively the same behavior as the one described in this letter.

*I-V* characteristics of junctions were obtained at 4.2 K by using a standard four-probe method with sweeping current while monitoring voltage. Fig. 2a shows the *I-V* curve of a bicrystal junction at zero field. Clearly, the junction displays a robust Josephson coupling between the two single crystals. The $T_C$ of the junction was found to be about 17 K and the *I-V* characteristic is resistively-shunted junction (RSJ) like with no obvious hysteresis. At 4.2 K, the critical current was about 660 μA, and the $I_C R_N$ product of the junctions had a value of about 10 μV. As shown in Fig. 2b, microwave induced Shapiro steps are clearly observed at voltages corresponding to multiples of *hf*/2*e* for the applied frequencies *f* of 2.5 GHz and 4.0 GHz. This confirms that the observed critical current is indeed a flow of Cooper pairs.

Fraunhofer-like diffraction patterns were observed for these bicrystal junctions by applying an external magnetic field parallel to the *a-b* surfaces at 4.2 K. Field modulations for positive and negative critical currents were found to be relatively symmetric, and a modulation period of about 2 gauss can be estimated from the pattern (Fig. 3). Taking the penetration depth of the iron pnictide ($\lambda \sim 200$ nm)[14] into account, the dimension of the effective coupling area ($W$), through which the Josephson current is flowing, is estimated to be about 25 μm for the bicrystal junction, consistent with the values obtained from our previous Pb/Ba$_{1-x}$K$_x$Fe$_2$As$_2$ junctions.[1] As discussed previously,[1] the effective junction area where the current is flowing could be significantly smaller than the nominally intended junction area defined by the overlapping of the crystals because much of the crystal surface layer is oxidized and thus non-superconducting. In particular, preliminary low temperature STM experiments performed on Ba$_{1-x}$K$_x$Fe$_2$As$_2$ single crystals suggested significant degradation in superconductivity of the surface after a period of exposure in ambient atmosphere.[1,15] Furthermore, following a method described in Ref. 16, the Josephson penetration depth ($\lambda_J$) of the junction is estimated to be about 26 μm, which in turn indicates a $W/\lambda_J$ ratio of about 1.0 for the junction. The value of $W/\lambda_J$ provides a self-consistency check for the obtained diffraction pattern:[16,17] the larger the value of the ratio, the larger the degree of the self-field effect. For the present junction, the near unity $W/\lambda_J$ ratio is consistent with the obtained fairly symmetric diffraction pattern, suggesting that the junction is in the intermediate regime.[17]

Previous phase-sensitive Josephson junctions/SQUIDs involving *d*-wave[18,19] or *p*-wave[20] superconductors were fabricated on two interfaces of a single crystal or across a twin boundary. Because the pairing symmetries in such junctions are anisotropic, a π phase shift occurs at one of the two interfaces. In contrast, the proposed *s*±-wave pairing symmetry is highly isotropic, and thus previous phase-sensitive experimental designs, such as corner junctions,[18] are not expected to provide direct tests for this pairing symmetry. In order to generate 'plus'-phase dominated and 'minus'-phase dominated faces which can be utilized to construct π-phase generating junctions for *s*±-wave pairing symmetry, several junction geometries have been proposed.[2-4] One such proposal involves a bicrystal junction that consists of both electron-doped (e-doped) and hole-doped (h-doped) iron pnictide superconductors. If the pairing symmetry in iron pnictide superconductors is indeed *s*±-wave, the two single crystals in the *p-n* structure will possess opposite signs in their bulk phases once the global phase coherence is established by Josephson current through the junction interface. Connecting the two single crystals in the bicrystal *p-n* junction with a conventional *s*-wave superconductor in a SQUID loop configuration[2] is expected to yield a π-shift at zero flux bias. The present demonstration of phase-coherent all pnictide *p-n* junctions opens the door to carrying out this phase-sensitive experiment.

In traditional semiconductor *p-n* junctions, diffusion of charges at the interface can lead to formation of a depletion layer, which is essentially responsible for the rectifying *I-V* characteristics. It has been suggested that such redistribution of charges may also take place in $YBa_2Cu_3O_7$/$Nd_{1-x}Ce_xCu_2O_4$ Josephson junctions[9] with an estimated deletion width of <1 nm.[21] In the case of iron pnictide *p-n* junctions, redistribution of charges could possibly lead to suppression of order parameters near the interface for both single crystals, which may play a role in the junction formation. However, because of the metallic nature of the parent compounds of the superconductors, the standard semiconductor depletion layer model may not apply to the iron pnictide *p-n* junctions, making the transport properties across the interface more difficult to analyze. As a characteristic signature of a junction with a depletion layer, asymmetric *I-V* characteristics over a large bias voltage range are commonly seen in semiconductor *p-n* junctions due to rectification. The present iron pnictide bicrystal junctions show highly symmetric *I-V* curves up to a bias voltage of ~100 µV. Because of the low values of the junction resistance, it is difficult to study high voltage bias characteristics with the present *p-n* junctions. In the future, one can explore fabricating higher resistance junctions using thin films and small junction fabrication techniques.

In summary, the Josephson effect has been observed in $SrFe_{1.74}Co_{0.26}As_2$/ $Ba_{0.23}K_{0.77}Fe_2As_2$ bicrystal junctions aligned along the *c*-axis of the single crystals. The *I-V* characteristics of the junctions are RSJ-like. The Fraunhofer-like magnetic diffraction pattern indicates that the junction is in the intermediate junction limit and the width of the active coupling area is of the order of 10's of micrometers. The success in establishing phase coherence between e-doped and p-doped iron pnictide superconductors in a bicrystal structure represents an important step in carrying out phase-sensitive measurements for probing the proposed $s\pm$-wave pairing symmetry. In addition, the present work also demonstrates the feasibility of fabricating all-pnictide Josephson devices.

The authors are grateful to I. I. Mazin, D. Parker and F. Wellstood for valuable discussions. X. Z. would like to thank K. Jin and P. Bach for technical help and useful discussions. Measurements at UMD were supported by the NSF under DMR-0653535; NPB is supported by the CNAM Glover fellowship; IT is funded by NSF MRSEC at UMD (DMR-0520471); and the work at SNU was supported by NRL (M10600000238) and GPP (K20702020014-07E0200-01410) programs.

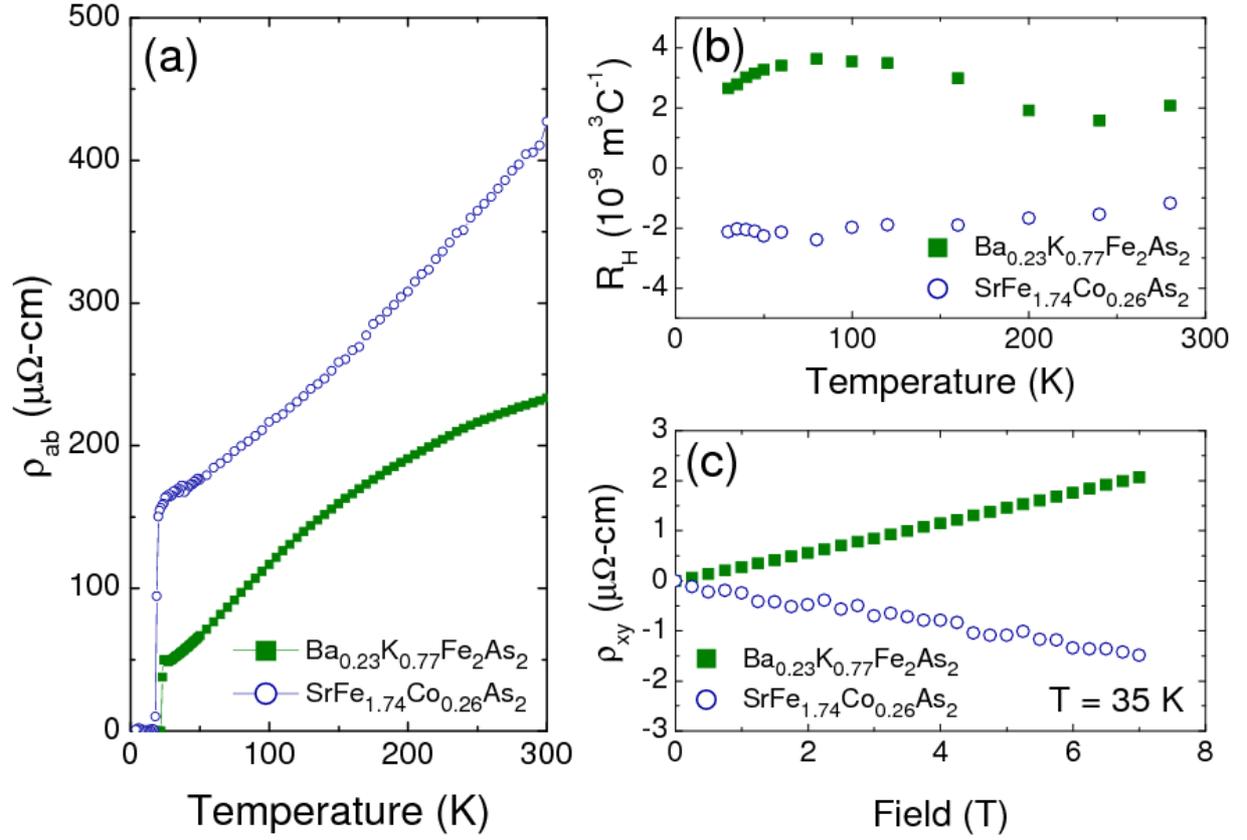

FIG. 1. (color online) Temperature dependence of the in-plane resistivity (a) and the Hall coefficient (b) for SrFe$_{1.74}$Co$_{0.26}$As$_2$ and Ba$_{0.4}$K$_{0.6}$Fe$_2$As$_2$ single crystals; (c) Field dependence of Hall resistivities obtained at 35 K for the two single crystals.

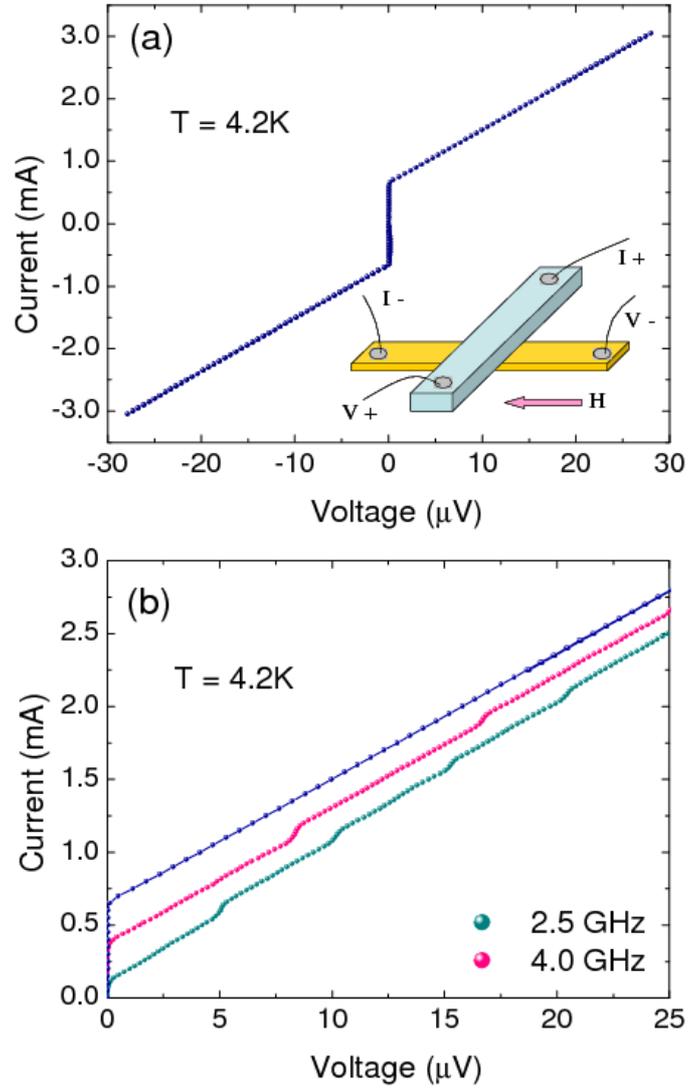

FIG. 2. (color online) (a) *I-V* characteristic of a bicrystal junction; the inset is a schematic view of the vertically aligned bicrystal junction structure with $SrFe_{1.74}Co_{0.26}As_2$ and $Ba_{0.23}K_{0.77}Fe_2As_2$ single crystals; the external magnetic field was applied parallel to the interface; (b) *I-V* characteristics of the bicrystal junction with and without microwave irradiation at two frequencies (2.5 GHz and 4.0 GHz).

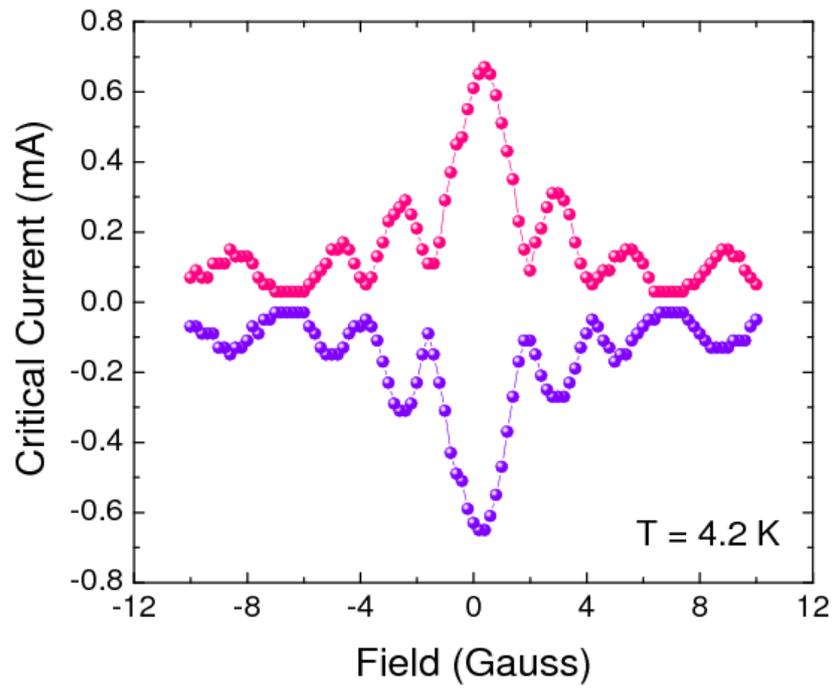

FIG. 3. (color online) Fraunhofer-like magnetic diffraction pattern of a bicrystal junction.